\newcommand{\CLASSINPUTtoptextmargin}{1.4cm}
\newcommand{\CLASSINPUTbottomtextmargin}{1.4cm}
\documentclass[conference]{IEEEtran}

\usepackage{graphicx}
\usepackage{enumitem}
\usepackage{amsmath}
\usepackage{subfig}
\usepackage{breqn}
\usepackage{pbox}
\usepackage{multirow}
\usepackage{url}
\usepackage{nohyperref}
\usepackage{flushend}
\usepackage{color}
\usepackage{listings}
\definecolor{dkgreen}{rgb}{0,0.6,0}
\definecolor{gray}{rgb}{0.5,0.5,0.5}
\definecolor{mauve}{rgb}{0.58,0,0.82}

\begin{document}

\title{Software Enabled Security Architecture for Counteracting Attacks in Control Systems }

%
\author{\IEEEauthorblockN{Uday Tupakula,
Vijay Varadharajan,  
Kallol Krishna Karmakar}
\IEEEauthorblockA{Advanced Cyber Security Engineering Research Centre\\
The University of Newcastle,
Newcastle, Australia}
\IEEEauthorblockA{[uday.tupakula, vijay.varadharajan, kallolkrishna.karmakar]@newcastle.edu.au}}

\maketitle

\begin{abstract}
Increasingly Industrial Control Systems (ICS) systems are being connected to the Internet to minimise the operational costs and provide additional flexibility. These control systems such as the ones used in power grids, manufacturing and utilities operate continually and have long lifespans measured in decades rather than years as in the case of IT systems. Such industrial control systems require uninterrupted and safe operation. However, they can be vulnerable to a variety of attacks, as successful attacks on critical control infrastructures could have devastating consequences to the safety of human lives as well as a nation’s security and prosperity. Furthermore, there can be a range of attacks that can target ICS and it is not easy to secure these systems against all known attacks let alone unknown ones. In this paper, we propose a software enabled security architecture using Software Defined Networking (SDN) and Network Function Virtualisation (NFV) that can enhance the capability to secure industrial control systems. We have designed such an SDN/NFV enabled security architecture and developed a Control System Security Application (CSSA) in SDN Controller for enhancing security in ICS against certain specific attacks namely denial of service attacks, from unpatched vulnerable control system components and securing the communication flows from the legacy devices that do not support any security functionality. In this paper, we discuss the  prototype implementation of the proposed architecture and the results obtained from our analysis. \\
\end{abstract}


\IEEEpeerreviewmaketitle


\section{Introduction}

Control systems are usually composed of a set of networked agents, consisting of: sensors, actuators, control processing units, and communication devices. Control systems such as electricity generation and supply, gas supply, logistics, manufacturing and hospitals are considered critical national infrastructure. The operational technologies that support critical infrastructure depend heavily on information systems for their monitoring and control. Traditionally, physical separation of ICS networks  was used as primary means for securing critical control systems.

Modern control system architectures, business requirements, and cost control measures result in
increasing integration of corporate and ICS IT architectures. Physical separation alone no longer provides
a viable business option for managing, utilizing, or securing ICS. The evolution of these systems from isolated environments into internet connected ones, in combination with their long service life and real-time nature have raised severe security concerns in the event of a cyber-attack. As a vital part of critical
national infrastructure, protecting ICS from cyber threats has
become a high priority since their compromise can result
in a myriad of different problems, from service disruptions
and economical loss, to jeopardising natural ecosystems and
putting human lives at risk. Stuxnet, Duqu, NotPetya,
and more recently, WannaCry, exemplify the devastating consequences this type of attack may have on critical ICS infrastructures. Hence there is urgent need for techniques that can be used to enhance the security of the control systems from cybersecueity attacks.  \\
The main aim of this paper to develop a software enabled architecture for securing control systems from cyberattacks. Our architecture uses software defined networks (SDN) and network function virtualisation (NFV) based techniques for enhancing security in control systems. SDN enables programmable networks and provides separation of control and data planes, enabling management protocols to be separated from the data traffic. The control plane consists of a logically centralised controller, which can be distributed in practice, and has native applications for management of network devices. Data plane consists of both physical and virtual network devices and implement protocols for forwarding traffic based on flow rules as well as protocols for (e.g. OpenFlow) for communicating with the control plane. NFV allows softwarisation of network functions enabling the provision network services as Virtual Network Functions (VNFs). Hence NFV enables the critical network operators to deploy security services in a dynamic fashion.\\
Such a software enabled architecture enables secure dynamic policy-based decision making which can be particularly suitable for securing control systems. Control systems are autonomous decision-making agents making real-time decisions. Real time decision making requires the capability to make dynamic decisions to ensure higher availability of services which is an important requirement in control systems. The software defined architecture proposed in this paper is able to define fine grained flow-based security ensuring the availability of critical control subsystems. Another major challenge in control systems is the inability to make updates to certain critical components which in turn can make them vulnerable. The proposed architecture can make such unpatched control components to be accessed in a secure manner, only via secure flows from authorised users and devices, thereby protecting from attacks.  Furthermore, traditionally control systems are designed to operate in closed environments, which meant that often they do not possess much (or any) security functionality. However, with the increasing use of Internet protocols such as IP for the management of control system components (such as controllers), they are susceptible to various cyberattacks. The proposed software enabled security architecture provides the capability to secure the flows from these legacy devices that do not support the required security functionalities.\\

The paper is organised as follows. Section \ref{sec:AM} presents an overview of the SDN architecture and the challenges and requirements for enhancing security in control systems. In Section \ref{sec:pbsa} we propose software enabled security architecture for control systems that is based on SDN and NFV technologies. We first present a high level overview of the CSSA and then describe the important components of CSSA. We also discuss the specific security functions that are invoked in the switches to deal with the attacks from end hosts.  In Section \ref{sec:dis} we present some general discussion related to our software architecture.  In Section \ref{sec:imp} we present the implementation of CSSA and demonstrate its usage by considering an example attack case scenario and how CSSA can help to detect such attacks. Then we present some performance results. Section \ref{sec:rel} considers some related work and Section \ref{sec:con} concludes the paper.

\section{SDN Overview and Security Requirements } \label{sec:AM}
In this Section, we provide an overview of the SDN Architecture and discuss some of the requirements for securing control systems from cybersecurity attacks. 
\subsection{SDN Overview:} 
\begin{figure}
    \centering
    \includegraphics[scale=.4]{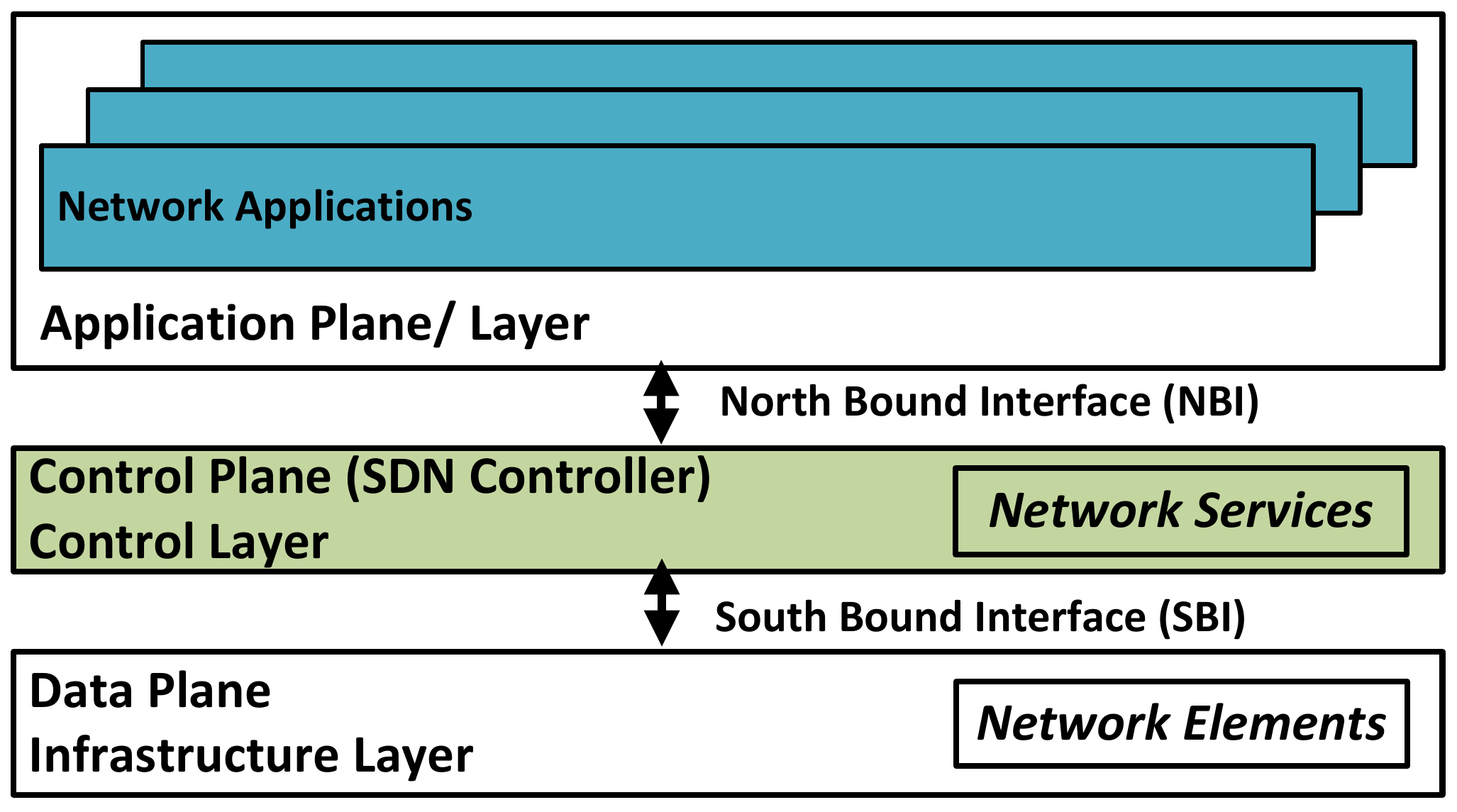}
    \caption{Software-Defined Network Architecture}
    \label{fig1:SDN}
\end{figure}
The SDN is a new architecture that has been deployed to enable more agile and cost-effective networks. It is an architectural approach that optimizes and simplifies network operations by more closely binding interactions such as provisioning, messaging, and alarming among applications, network services and devices, whether they be real or virtualized \cite{nadeau13}. The key features of SDN are \cite{nadeau13}:
\begin{itemize}
\item The separation of the network control plane (SDN Controller) from the data plane (Southbound Interface).
\item A logically centralized controller communicating with the data plane over open and standardized interfaces and protocols (OpenFlow).
\item The control applications (Northbound Interface/Programmable API interface) running on top of the Controller. 
\end{itemize}
Figure~\ref{fig1:SDN} shows the SDN architecture proposed by Open Networking Foundation~\cite{onf}. The SDN architecture  differs from legacy solutions by building networks  in three abstractions/planes/layers.
\begin{description}
\item[Data Plane/Infrastructure Layer. ] The Data Plane or Infrastructure Layer acts as the foundation for an SDN architecture. This plane consists of both physical and virtual network devices such as switches, routers, wireless access points. These devices implement the OpenFlow protocol to maintain communication with the Control Plane and also implement standard methods for forwarding traffic using flow rules.
\item[Control Plane/Layer. ] The Control Plane acts as the brain for the whole networking ecosystem. The Control Plane is decoupled from the underlying Data Plane infrastructure to provide a single centralised view of the entire network.  The Control Plane utilises OpenFlow to communicate with the Data Plane devices. The SDN Controller is a logically separate entity responsible for receiving instructions from the application layer and enforcing them over the Data Plane devices. It is also responsible for gathering information about network devices, events and statistics and sharing this data with the network applications running in the Application Plane.
\item[Application Plane/Layer. ] The Application Plane consists of network services, orchestration tools, and business applications that interact with the control layer. For instance, an application to monitor malicious activity in Data Plane devices. These applications leverage open interfaces to communicate with the Control Plane and the network state. 
\item [API interfaces. ] The application programming interfaces (APIs) are an alternate way to provide the abstraction necessary for SDN, along with a highly programmable infrastructure. APIs provide a channel by which instructions can be sent to a device to program it. In SDN, APIs are called \lq\lq Northbound\rq\rq or \lq\lq Southbound\rq\rq , depending on where they function in the architecture. APIs that reside on a Controller and are used by applications to send instructions to the Controller are northbound because the communication takes place north of the Controller. Southbound APIs reside on network devices such as switches. These are used by the SDN Controller to provision the network, with the communication taking place south of the Controller.
\end{description}
\subsection{Security Requirements for Control Systems}
\begin{itemize}
\item Control systems have higher availability requirements compared to systems in enterprise networks.  Control systems are autonomous decision making agents which need to make decisions in real time. While availability is a well studied problem in information security, real-time availability provides a stricter operational environment than most traditional IT systems. Hence there is a need for techniques that can make real time decisions for higher  availability of services in the control systems. The number of connections that can be handled by the servers depend on the application or hardware requirements. In our model, the SDN Controller is used to limit the number of flows to server depending on the server capacity. The flow control can be enforced on a per flow, per device, per domain or per location.
\item The property of control systems that is most commonly brought up as a distinction with IT security is that software patching and frequent updates, are not well suited for control systems. For example, upgrading a system may require months of advance in planning of how to take the system ofﬂine; it is, therefore, economically difﬁcult to justify suspending the operation of an industrial computer on a regular basis to install new security patches. Some security patches may even violate the certiﬁcation of control systems. For instance, as discussed in \cite{krebs2008cyber}, a nuclear power plant was accidentally shutdown because a computer used to monitor chemical and diagnostic data from the plant’s business network rebooted after a software update. When the computer rebooted, it reset the data on the control system, causing safety systems to incorrectly interpret the lack of data as a drop in water reservoirs that cool the plant’s radioactive nuclear fuel rods. Hence there is a need to ensure accessibility to unpatched system while protecting it against the attacks. In our approach, SDN is used to provide access to the authorised users and the flows from the authorised devices are protected against attacks.
\item ICS protocols were initially designed to operate in closed
environments. Authentication and encryption of the communication between the devices was not considered as requirement during initial design. The lack of security features in ICS protocols remained
largely unnoticed due to the deployment in isolated (trusted)
environments. This changed recently when ICS protocols have
been stacked onto IP, enabling the management of ICS controllers via the global Internet. Such communication requires
protective measures, either via secure tunnels between trusted
domains or end-to-end authentication and encryption. Unencrypted ICS traffic is particularly dangerous since it is
prone to eavesdropping and manipulation attacks. Hence there is a need for techniques for securing the flows from the leagcy and resource constrained devices that do not support any security functionality or protocols. 

\end{itemize}


\section{Control System Security Application} \label{sec:pbsa}
\noindent In this section, we will first discuss some of the assumptions and overview of our model. Then we  describe the CSSA logical architecture and specific security functions developed for switches to deal with the attacks in control systems.  
\begin{itemize}
\item We assume that the Controller in the SDN domain is not compromised.  This is a common assumption for security models in SDN. 
\item We use the term end host to represent a client or server or field device. We assume that end hosts are connected using SDN switches. Hence any communication between the end hosts pass through the switches. 
\item A flow is used to represent communication between the end hosts. A flow can be a single packet or sequence of packets between the end hosts. 
\end{itemize}
  
\subsection{Overview:} 

\begin{figure}
    \centering
    \includegraphics[scale=.8]{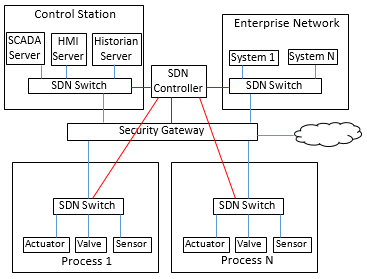}
    \caption{Sample Scenario}
    \label{contrl1}
\end{figure}

We consider a simple scenario as shown in Figure \ref{contrl1} where the control station with different servers such as SCADA server, HMI server, Historian server is used to monitor several processes with various filed devices such as actuators, valves and sensors in a control system. The enterprise network is also connected to the control system network using the security gateway. The field devices, servers in the control station and the enterprise systems are interconnected using SDN switches and the SDN controller is used for managing the complete network.

\begin{figure}
    \centering
    \includegraphics[scale=.9]{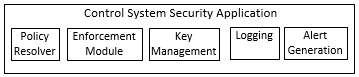}
    \caption{ Control System Security Application Components}
	\label{CSSA}
\end{figure}

Figure~\ref{CSSA} shows the important component of the CSSA which is designed to run as an application on SDN Controller. 
 We have used a modular approach in the design of CSSA and NFV enables implementation of specific security functions depending on the capability of switches. The CSSA is used in the specification and evaluation of security polices for managing all the network devices in a SDN based control system domain. The SDN Controller has global knowledge of all the devices in the network. Since the CSSA is implemented on the SDN Controller, it is able to access the information available in the Controller and use it for monitoring and detecting the attacks from the end hosts. Now let us consider the high level operation of the CSSA. \\
 When the switches receive a new flow, they generate a Packet\_in message to the SDN Controller which is subsequently forwarded to the CSSA. The CSSA retrieves the specific policies related to the flow and invokes specific security related functions for establishing secure communication path and monitoring the flows for attacks.\\  
 Such decisions include establishment of routes in the SDN that meet the requirements of the end users/applications and whether flow requests from the end hosts are to be permitted according to the security policies as well as generation of keys for establishing secure routes and communications between the end hosts. We use OpenFlow protocol for communications between the SDN Controller ad the switches. Also, we have used  XML for specification and communication of security policies between the CSSA and the switches.\\ 
The main advantage of enforcing the security policies at the switches is that the attack can be detected and prevented nearest to the malicious end host (from where it comes from).  
  
\subsection{Control System Security Application}

The CSSA has security policies for all network devices in a given SDN based Control System domain. The CSSA enables the security administrator(s) to specify policies based on different parameters and conditions such as time, location, source, destination, event and type of traffic. For instance, policies can specify access to services during specific time intervals and/or using certain devices from specific locations. Similarly there can be different policies depending on the source address and/or destination address.  Also, there can be different policies that can be triggered during different events. For instance, when there is a legitimate increase in the traffic load, there can be mechanisms to replicate the services on multiple servers and distribute the load between them; during off-peak times, when the traffic falls below the certain lower threshold, then the services hosted on multiple servers can be contracted to a few servers. However, if the increase in the load is caused by an attack, then the policies will create mechanisms to isolate the hosts that are generating the attack traffic. There can also be security policies for different types of traffic allowing secure end to end paths for sensitive communications between the end hosts and field devices in sdn managed control system domain.\\
The main components of the CSSA are: Policy Resolver, Key Management, Enforcement module, Logging and Alert Generation. 
Policy Resolver is used for storing the policies related to the domain and determining the appropriate policies. The Enforcement Module derives the mechanisms for enforcing the policies determined by the Policy Resolver. The Key Management module is used for generating keys for secure communications. Logging module is used for logging all the messages transferred between the Controller and the CSSA and Alert Generation is used to raise alert to the Security Administrator when attacks are detected in the control system network. \\
The CSSA receives the packet\_in message from the SDN Controller and forwards it to the Policy Resolver to determine the related security policy. The Policy Resolver analyses the packet\_in message (such as the source address and destination address) and retrieves the related security policy from its database and updates the CSSA with the related policy. The Enforcement Module is used for enforcement of the policies locally or dynamically invoking the specific functions at the switches in the data plane.  If a new security policy mandates secure communications, then the CSSA requests the Key Management module to generate keys for secure communication and distributes the key to the corresponding switches (connected to the respective end hosts).The alert generation component generates an alert to the security administrator when attacks are detected in the control system network. The administrator can conduct an offline analysis to determine the seriousness of the event and decide to either isolate the suspicious end host from the network or restrict the end host communication depending on the availability requirements of the end host.
\subsection{Switch Security Functions}
In this section we will describe the important security functions that are implemented at the switches. 
\subsubsection{Logical Store()}
This LS()function is used for storing specific information about the end hosts which are useful for detecting attacks. For instance, LS() is used to store  attack signatures, whitelists, blacklists,  behaviour profiles of the end hosts and audit records related to the incoming and outgoing traffic from the end hosts. 

\subsubsection{Traffic Validation()}

The TV() is used in the validation of the traffic from all the end hosts connected to the switches.  It makes use of source address validation, signature and anomaly based techniques for detecting attacks. Validating the source address prevents an end host from injecting malicious messages and generating attack traffic with spoofed identities. The signature and anomaly based techniques are then used to deal with the attacks that are generated with correct source address. 

We have developed software modules for implementing deep packet inspection based validation in the OpenFlow switches. The software modules analyse the payload and apply access control mechanisms to drop the malicious packets. Access control rule matching is done using Regular Expressions. GNU C regex has been used for this purpose. As an Open vSwitch is built mostly using C, it makes GNU C regex easily adaptable to the environment. Comparison of the access control rule with the payload information is achieved by storing the rules in LS() and then developing a regular expression string for each access rule. Then the created regex strings are compared with the payload information. TV() accepts the access control rules sent by the CSSA and stores them in an array. When a packet flow occurs, the TV() pulls the corresponding access rules and creates regex strings, which are then matched against the packet payload. If a match is found, TV() drops the flow and raises an alert to the CSSA. \\
The anomaly detection mechanisms use static thresholds at the initial stages to prevent flooding attacks on critical services in the control system. The threshold is determined by the security administrator based on the critical server application, OS and hardware used for the critical services. This thresholds is used to validate the flow requests to the server in real time and limit the flows based on different factors such as per device, per switch, per domain and per location.  
Traffic logs captured at LS() are used for training the TV() to capture the network behaviour. As part of our future work, we will develop  machine learning  for capturing the behaviour of the network traffic and detecting attacks. The trained classifier is stored as a baseline detector for matching the traffic behaviour. During testing, TV() uses these pre-compiled profiles for validating the behaviour of the end hosts and detecting the attacks. 

\subsubsection{Flow Encryption ()}
The FE component is used for securing the communications between the end hosts. If the flow security policy mandates secure communication between the end hosts, the key management module in the CSSA will generate the required symmetric key for securing the flows and distribute the key to the relevant switches. \\
In control systems there can be several legacy and resource constrained devices that do not support any security functionality or protocols. Hence, in these cases, security administrators can make use of this FE() for securing the communication between the hosts. If a new flow is destined to devices/hosts that do not support security functionality, then the CSSA can enforce policy to encrypt the flow at the switch that is connected to the source host and decrypt the flow at the switch that is connected to the destination host.

\section{Discussion}\label{sec:dis}
In this section, we will discuss how our model can enhance the security  in control systems.\\
As the Controller has the visibility of its network domain topology and devices, the CSSA uses this information to make real time decisions on communication between the end hosts based on their availability requirements. The traffic flows initiated by the end hosts are subjected to security policies in the CSSA in the SDN Controller of that domain. The source host could be any client machine or field device. The initial packet header from the source host is sent by the switch (to which this host is connected) to the SDN Controller which is subsequently forwarded to the CSSA.  The header contains all the usual network and service parameters such as the source address, destination address the packet type. The CSSA extracts the relevant parameters from the incoming packets and uses the Policy Resolver to determine whether the communication or flow is permitted according to the policies configured by the security administrator.  If the flow is permitted by any of the polices, then the path is established to enable communication between the end hosts. If the flow is not permitted by any policy then the flow is dropped. If the end hosts still continues to generate excessive flow requests then an alert is raised to the security administrator. This
is a significant advantage as it helps to deal with attacks such as outbreak of worms in critical control systems. For instance, the Slammer worm which randomly
scanned for vulnerable machines for spreading the attack created severe congestion in the network and as well as  disabling safety monitoring system in nuclear power plant for over five hours \cite{poulsen2003slammer}. Using the proposed architecture, CSSA drops all the
randomly generated malicious flow requests which do not satisfy policies permitting the flows. As the malicious flows are dropped at the source end, the impact on the bandwidth is minimized. This can be significant in certain control systems.\\
Case 1: Now assume that there is a specific server providing a critical service and consider how CSSA can help to achieve higher availability for that critical service. Assume that some end hosts in the control system are compromised by an attacker and used to generate flooding attacks on the server that is hosting the critical service. Without the proposed architecture, the attacker can flood the critical service severely impacting the availability of the service. Using the proposed architecture, the CSSA uses the network domain visibility to make decisions, in real time, on communications between the end hosts based on their availability requirements. If the critical server is capable of handling x flows/sec, then the CSSA  restricts the flows to the critical server to less than x flows/sec. The dynamic policy driven approach in the CSSA allows the flows to be restricted based on a variety of parameters such as maximum number of flows per end host, maximum number of flows per switch and maximum number of flow per control process or logical domain. \\
Case 2: Protecting unpatched server. In this case the flows to the unpatched server can be permitted only from selective hosts. Furthermore, all the flows from the permitted hosts can be monitored for attack traffic and the malicious end hosts can be isolated if they are attempting to exploit the vulnerability of the unpatched server. In the following section we will show how the end host malicious flow is blocked from exploiting the Shellshock vulnerability on the unpatched web server. \\
Case 3: The ICS processes have a wide range of time-related requirements, including very high speed, consistency, regularity, and synchronization.  Some systems may require the computation to be performed as close to the sensor and actuators as possible
to reduce communication latency and perform necessary control actions on time. It is important to meet these time-related requirements when designing security for the ICS. In our model, the CSSA provides a guaranteed path between the specific sensors and actuators to meet the time related requirements for the safe operation of control systems.

\section{Implementation}\label{sec:imp}
\begin{figure}[!ht]
    \centering
    \includegraphics[scale=.35]{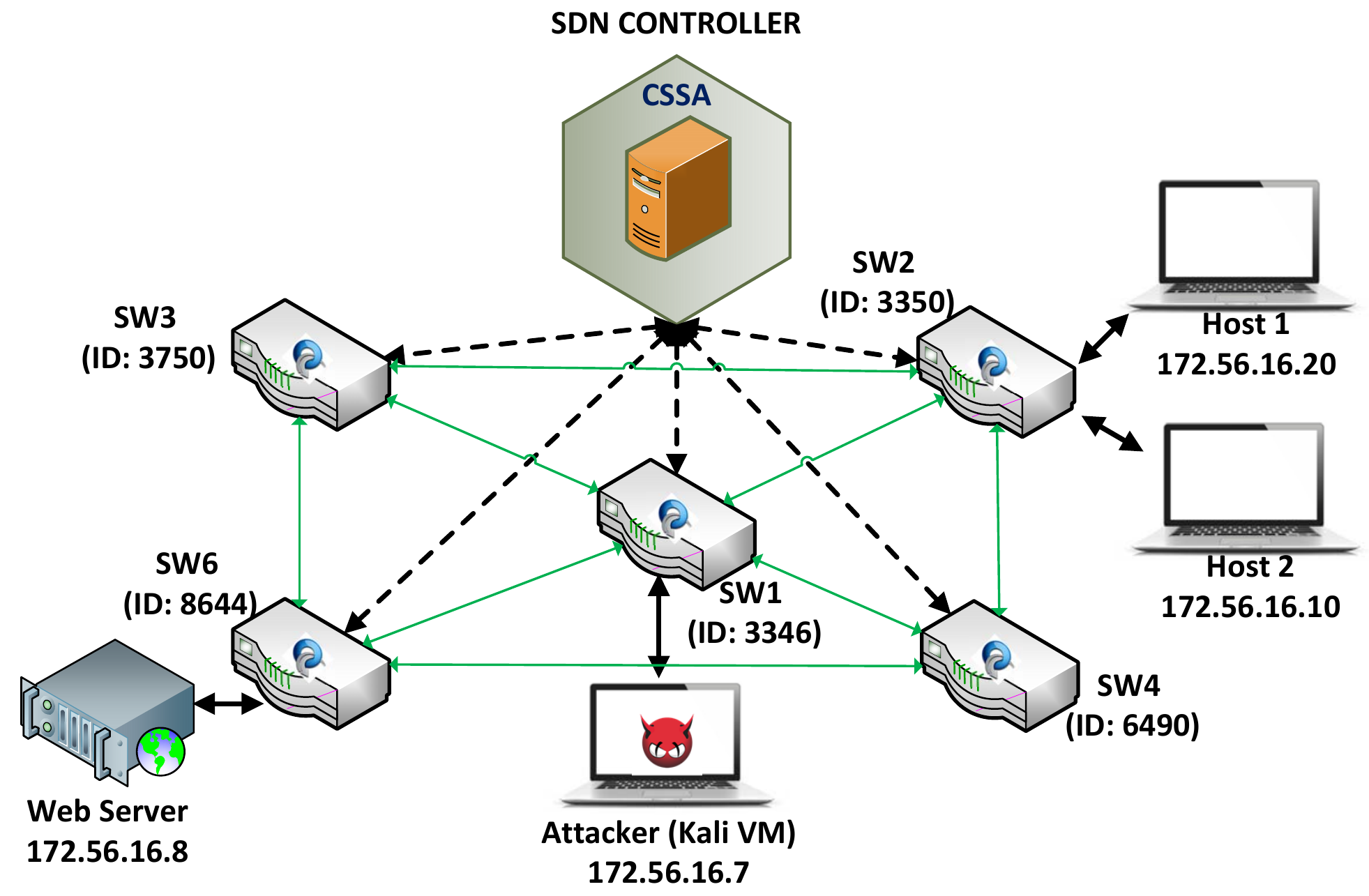}
    \caption{Network Setup}
	\label{network}
\end{figure}

We have used ONOS as the SDN Controller and Open Virtual Switches to justify our CSSA. The machine 
we are using for simulation is a Dell Power Edge M640 Server, consisting of two Intel Xeon Gold 6126 @ 2.6 GHz processors and 384 GB(6x64) of RAM. We have used Oracle VM Box and open virtual switches to create the network environment. Our network configuration is shown in Figure~\ref{network}.\\ 
We have developed the CSSA as an ONOS application running over North Bound Interface. Several modules in ONOS such as switch manager, module management, link discovery and REST APIs are adapted from the Floodlight Controller. The data model in ONOS is implemented using Titan graph database, Cassandra key value store and the Blueprints graph API to expose network state to applications. Applications read the global network view to make forwarding and policy decisions and write their policies to be enforced on the network view. \\
Open vSwitch (aka) (OVS) is an open-source multilayer virtual switch used for network virtualization. It helps to establish control over distributed physical servers using virtualised switching abstraction. It is implemented using C. However, it can be adapted  to various environments. It supports and uses OpenFlow. It also supports other protocol like NetFlow, and sFlow. In most cases, it is used in data centres. It is supported by most Linux oriented virtualization technologies including Xen/XenServer, KVM and Virtual Box.\\
The Host 1 is RTU/PLC slave machine (172.56.16.20) simulates
the acquisition of parameters from 10 boiler water tanks and a
main tank, and one valve actuator controlled from the PLC.
The PLC node is simulated using the Modbus PLC Simulator,
installed on a Windows 10 virtual machine. Host 2 (172.56.16.20) runs the SCADA MTU/PLC MASTER which is based on a free version of Promotic PLC running on a Microsoft Windows Virtual Machine. The MTU machine also runs the SCADA HMI to visualize the process. The web-server is based on DVWA, a XAMPP
web-server designed with vulnerabilities configurable with
difficulty from low to impossible which is hosted on linux OS and vulnerable to shellshock. The kali linux machine is used for generating attacks. Modbus-cli tool is used for  reading and writing values to the memory registers. Attackers can use this tool to capture the data from the target any analyse the data before performing the attack. For instance, attackers may want to know which coil controls the pressure valve on the system and what value that pressure value has in its register that is the threshold for opening and closing it. If these values are changed, the entire plant could be placed at risk.  \\

\subsubsection{Defending unpatched systems} 
There are an increasing number of zero day attacks \cite{istr16} each year. In general, it can take several weeks for the patches to be released by the vendors \cite{htbridge14}. Furthermore, even after the patches have been made available, an attacker may be able to exploit the vulnerabilities in practice, as systems may not have been patched \cite{grossman13}; this is a common occurrence in the real world. 

With critical infrastructures such as SCADA systems, network operators can be hesitant to apply the patches, as in some cases this can require the systems to be taken offline; also sometimes the operators can be worried that patching can break the SCADA application. As long as the systems remain unpatched, they are vulnerable to attacks. Our security solution helps to protect such vulnerable systems by preventing the malicious traffic at the switch that is near to the source of attack. 

\begin{figure}
\centering
\subfloat[]{{\includegraphics[scale=.45]{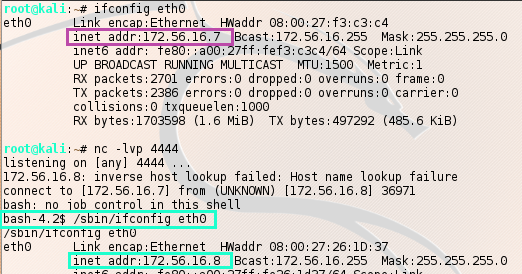}\label{shellshock}}}\\
\subfloat[]{{\includegraphics[scale=.3]{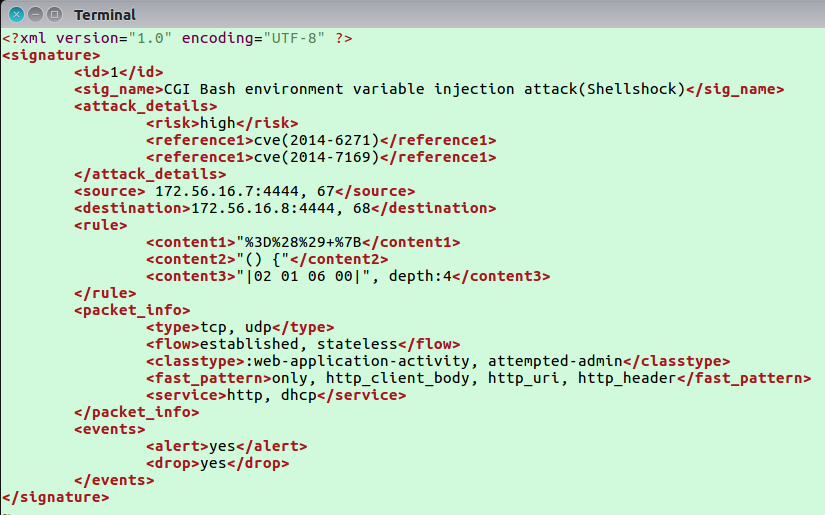}\label{shellshock1}}}\\
\subfloat[]{{\includegraphics[scale=.56]{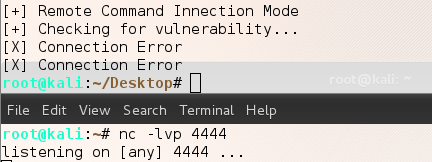}\label{pencon}}}
\caption{Reverse Shellshock attack (a) Successful approach (b) Attack Rule in XML specification and (c) Attack prevention at Switch connected to malicious source}
\label{shell}
\end{figure}

In Figure~\ref{network}, we have used a Kali Linux VM as the malicious client VM to launch a Shellshock attack on an unpatched Web Server VM. Figure~\ref{shellshock} shows a successful Reverse Shellshock attack on the vulnerable Web Server without our security solution. 


With our security solution, attack signatures shown in \ref{shellshock1} are applied at SW1 switch to prevent malicious client VMs from targeting attacks on the unpatched web server. Then when we launched the same attack on the Web Server VM, the attack was not successful and the malicious VM is isolated from the network. As shown in Figure~\ref{pencon}, the malicious VM1 lost its connectivity. 
	
In traditional networks, such attack detection and prevention are performed at the destination host. With our security solution, such attacks are detected and prevented {\em nearer to the source that is generating the malicious flow}. This is a significant advantage since it enables the resources at the destination and the switches to be available for legitimate traffic. That is, the impact of the attacks on the network resources and bandwidth is reduced. Furthermore, the compromised end host is isolated from the network to prevent further attacks on any other host. The security administrators are able to carry out offline analysis to determine how the attacker compromised the end host and take proper measures to update the system before re-connecting it to the network. 
Table \ref{tab:traf} presents the worst case latency results for the varying number of rules used for traffic validation using DPI. 

We have presented the average worst-case results for 10 runs for varying numbers of rules for a single flow. We reset the system after each run to ensure that the rules are flushed out, thereby preventing any caching of the rules. This is important because if we use multiple flows, the flows can be either dropped or forwarded when they match with any of the rules. Therefore, even if we have 1000 rules and if the incoming packet matches with the first rule, then we may not observe any delays. Hence, we have provided the results for the case of a single flow, which only matches with the last rule. To ensure that the flow only matches the last rule, we have used deny policies for all rules and used the permit only for the last rule. Hence, even if the packet matched any of the earlier rules, it will be dropped and we can detect the inconsistencies with the rule matching. So the results represent the actual worst case delays with our architecture components.

\begin{table}
\centering
\caption{Traffic Validation}
\label{tab:traf}
\begin{tabular}{ccc}
\hline
Rules & Delay \\ \hline
10 & 0.008s \\ 
50 & 0.0112s \\ 
100 & 0.0264s \\ \hline
\end{tabular}
\end{table}
 \subsubsection{Flow Encryption and Decryption} 
 We have added \textit{USB to Ethernet} converters to make switch ports. We extended the LINC-Switch \cite{lisw} to support encryption using \textit{crypto} module. We have used AES algorithm with key size 128 bits.
\begin{figure}
\centering
\includegraphics[scale=.6]{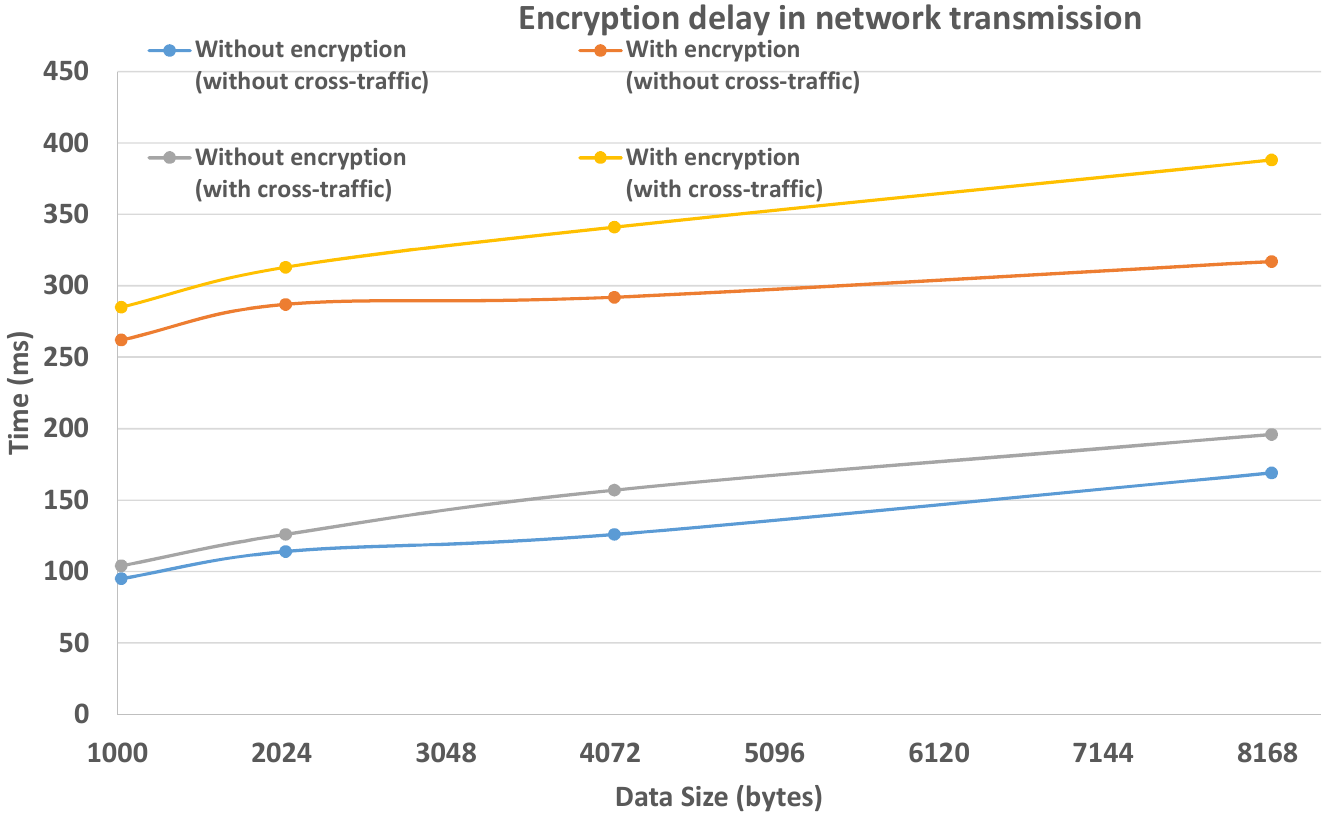}
\caption{Flow encryption-decryption time}
\label{fig:edtime}
\end{figure}
Figure~\ref{fig:edtime} shows the end to end delays caused due to the encryption and decryption of the flows for variable data sizes for the case scenario of 200 hosts connected to each edge switch with path length of four. The results are shown for the cases of flow without encryption without cross traffic, without encryption with cross traffic, with encryption and decryption without cross traffic and with encryption and decryption with cross traffic. As shown in the figure, the encryption and decryption of the flows incurs additional delays to the flow. The delay also increases with the increase in the data size and for the cases with cross traffic. However, these delays are applicable for the cases where there is need for dynamic encryption of flows such as legacy systems in critical infrastructures which may not support any security functionality. Hence it is of significant advantage for enhancing the security of the legacy applications.
\subsection{Performance}
In this section, we will present some of the performance results of CSSA running over ONOS. 
\begin{figure}[!ht]
    \centering
    \includegraphics[scale=.35]{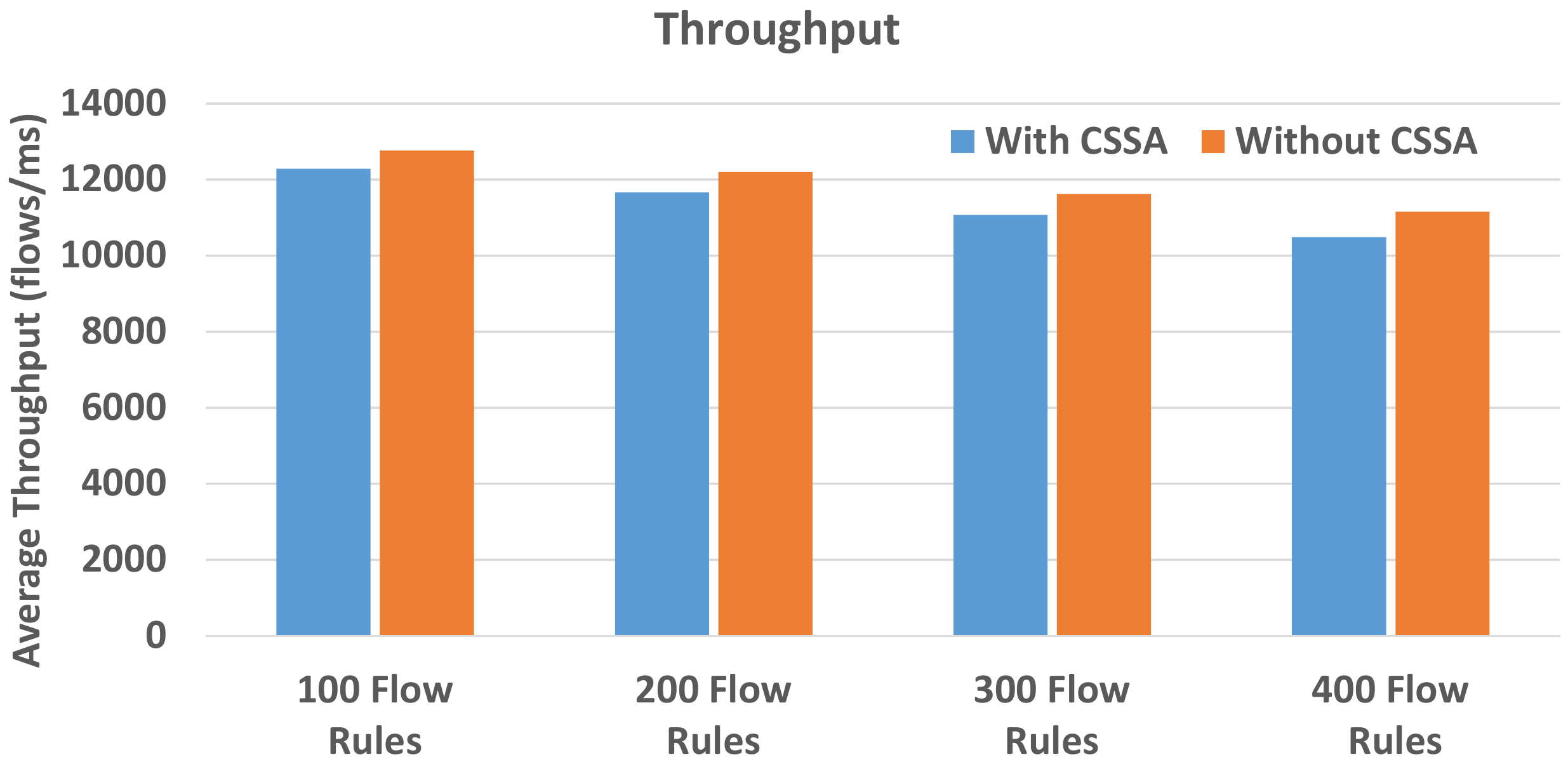}
    \caption{Throughput}
    \label{fig:throughput}
\end{figure}
\begin{figure}[!ht]
    \centering
    \includegraphics[scale=.5]{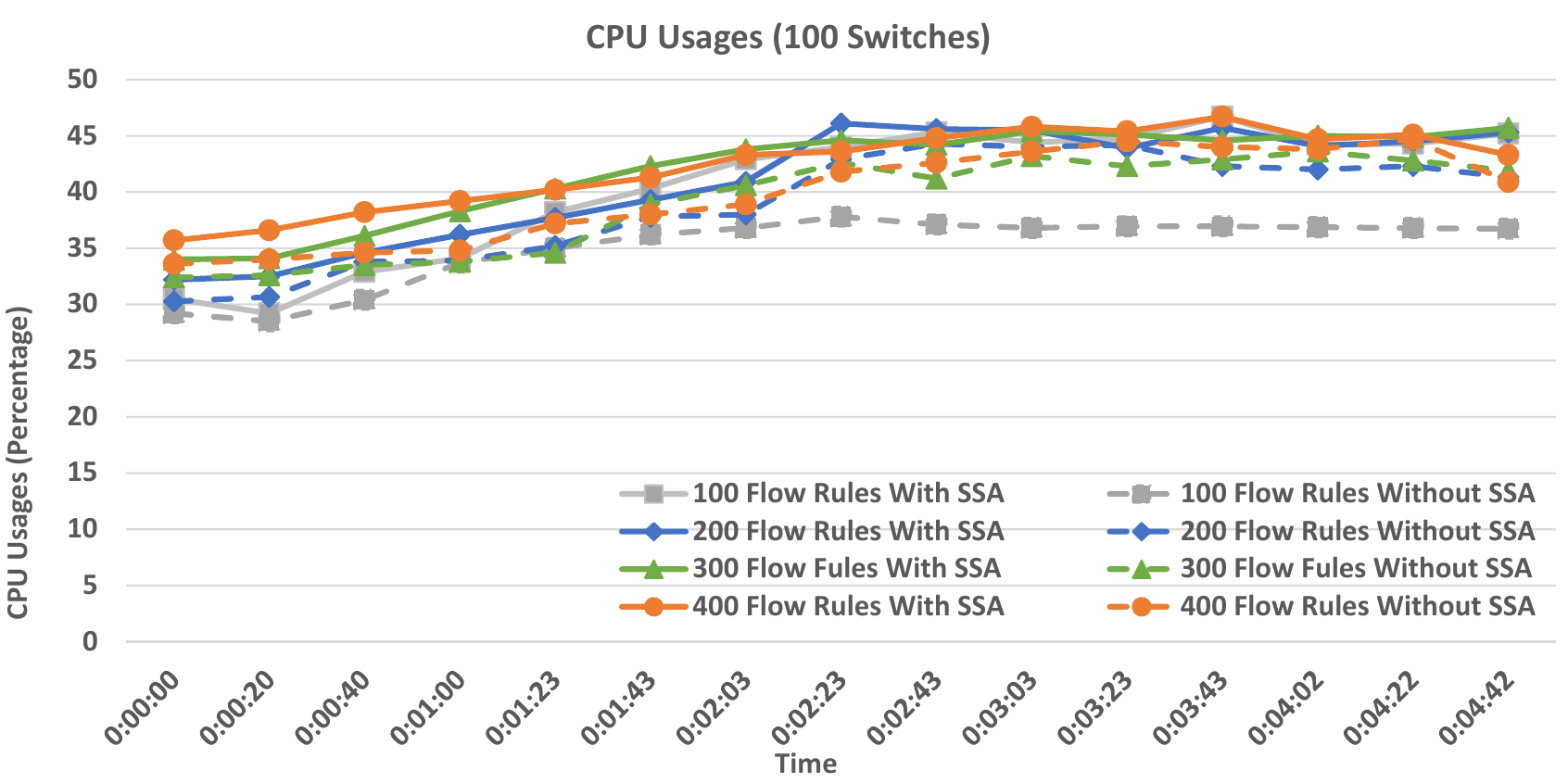}
    \caption{CPU Usages}
    \label{fig:cpu}
\end{figure}  
\begin{figure}[!ht]
    \centering
    \includegraphics[scale=.52]{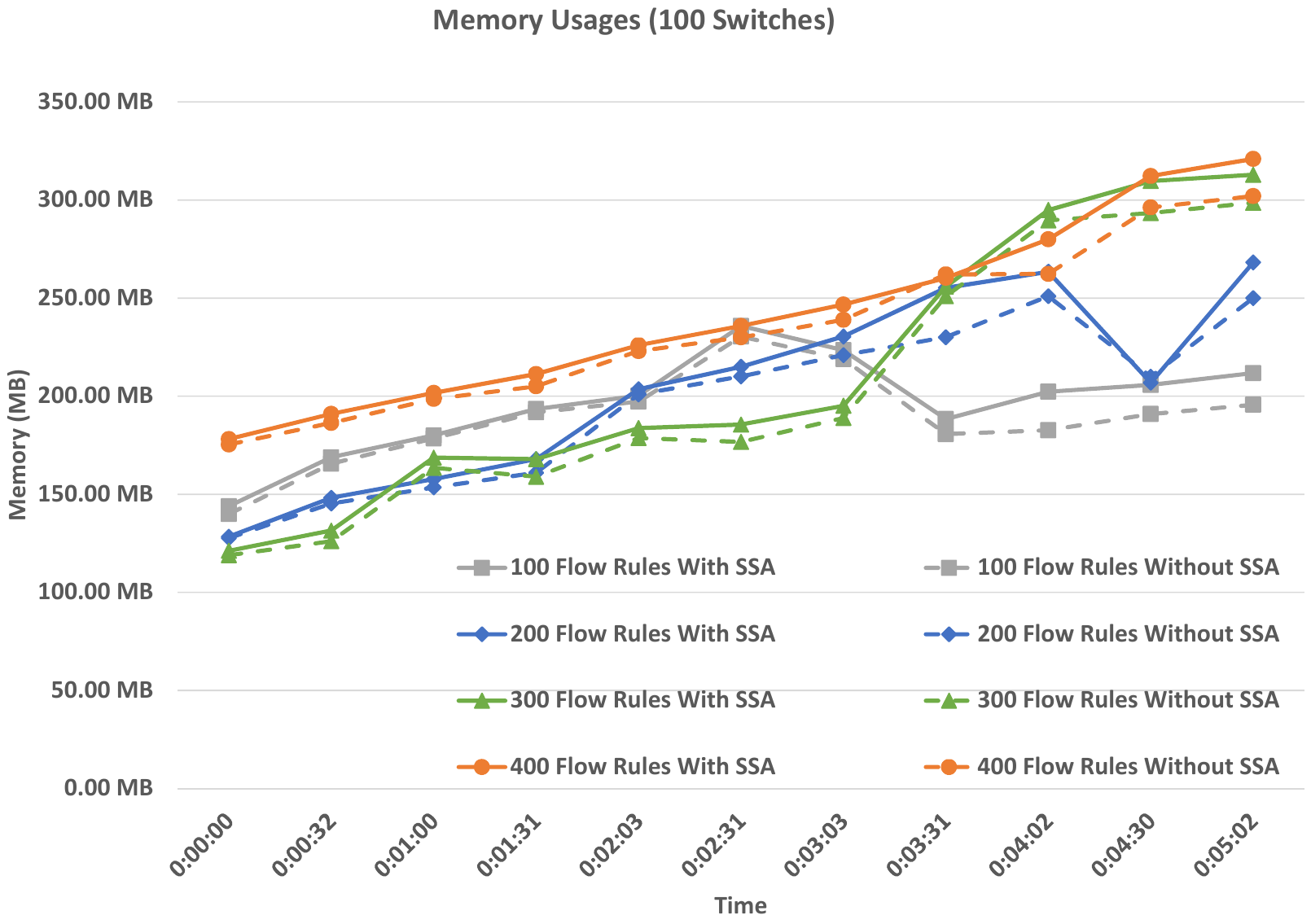}
    \caption{Heap Memory Usages }
    \label{fig:heap}
\end{figure}
We have used CBench, an SDN controller benchmarking application to measure the throughput.  CBench works by creating a set of switches and sending different types of traffic between the switches and the Controller. We have measured the flow rate with and without the CSSA for 20 switches where each switch is connected to 10 end hosts and the number of flow rules in the switches are increased from 100 flow rules to 400 flow rules. Figure \ref{fig:throughput} shows the throughput with and without CSSA. The orange bars shows the throughput obtained by activating default applications such as the ONOS core, drivers, proxy and the forwarding application. Blue bars show the throughput of the Controller after activating CSSA together with the previously mentioned applications. As shown in the Figure \ref{fig:throughput}, the throughput decreases with the CSSA and it also varies in accordance with the number of flow rules in the switches. Typically the reduction in throughput is between 4\% to 6\%.

With the similar setup, we have used Jconsole to measure the CPU usages and Heap memory usages while ONOS is working with and without CSSA. Jconsole is a built-in API of JAVA to monitor the individual JAVA modules running within the JVM. 
Figure \ref{fig:cpu} shows the CPU usages, and Figure \ref{fig:heap} shows the Heap memory usages without and with CSSA for varying number of flow rules. In both the cases, the overhead increases with the increase in the number of flow rules in the switches. 

\section{Related work}\label{sec:rel}
In this Section we will present some of the related work. 
McLaughlin \cite{mclaughlin2016cybersecurity} presented a detailed study on the threat landscape of ICS. Nawrocki \cite{nawrocki2019uncovering} conducted a detailed study on the security impact of connecting ICS to Internet. The traffic analysis is correlated with data from honeypots and Internet-wide scans to separate industrial from non-industrial ICS traffic and the authors provide an in-depth view on Internet-wide ICS
communication and identified several critical risks to the ICS and Internet.    \cite{barrere2019identifying} proposed security metric based on AND/OR graphs that represent cyber-physical dependencies among network components. The proposed metric is able to identify sets of critical cyberphysical components, with minimal cost for an attacker to make the control system into enter into a non-operational
state. Niedermaier \cite{niedermaier2019cort} developed testbed for testing ICS against specific attacks where an attacker has remote access to the network
and an attacker who has local access to the ICS components
with basic knowledge and \cite{shalyga2018anomaly, kim2019anomaly} proposed anomaly detection techniques for ICS. \\

Currently SDN is increasingly being deployed in different networks and there is ongoing work  ~\cite{sezer13we,   lee2017delta, dacier2017, marin2019depth} for identifying different possible attacks in such networks and developing techniques \cite{jiasi19secure, park19dpx, shin2014rosemary, porras12, li14, chaignon2018oko} to deal with the attacks in SDN. Erokhin \cite{erokhin2019critical} presented some preliminary study for using SDN to monitor the critical information infrastructure.  We have proposed a software enabled architecture that is based on SDN and NFV technologies. We have also described how our architecture is able to address the specific security requirements of the control systems. \\

\section{Concluding Remarks}\label{sec:con}
In this paper, we have proposed a software enabled security architecture for securing control systems. Our security architecture uses SDN technology to dynamic policy driven decision making across the network infrastructure. Such an approach enables dynamic isolation of control system components and network devices that are vulnerable to security attacks. The security architecture uses NFV technology to provision dynamically security functions at the required places in the network infrastructure thereby enhancing the security capability of control system components. We have used this combined SDN and NFV technologies enabled security architecture to demonstrate how it can be used to protect control systems from certain specific attacks namely denial of service attacks, from unpatched vulnerable control system components as well as securing the communication flows from the legacy devices that do not support any security functionality.\\


\bibliographystyle{IEEEtran}
\bibliography{ref}

\end{document}